\documentstyle[mncite,draft,psfig]{mn}

\title[The recurrent nova U Sco]
{The mass of the white dwarf in the recurrent nova U~Scorpii}

\author[T. D. Thoroughgood et al.]
{T. D. Thoroughgood,$^1$\thanks{E-mail: Tim.Thoroughgood@shef.ac.uk} 
V. S. Dhillon,$^1$ 
S. P. Littlefair,$^1$ 
T. R. Marsh,$^2$ 
D. A. Smith$^{1,3}$\\
$^1$Department of Physics and Astronomy, University of Sheffield, 
Sheffield S3 7RH, UK \\
$^2$Department of Physics and Astronomy, University of Southampton, 
Highfield, Southampton SO17 1BJ, UK \\
$^3$Winchester College, Winchester SO23 9LX, UK}

\date{\center{Accepted for publication in the Monthly 
Notices of the Royal Astronomical Society \\ 
\vspace{.5cm} 17/07/2001}} 

\begin{document}
\maketitle

\begin{abstract}
We present spectroscopy of the eclipsing recurrent nova U~Sco. The radial 
velocity semi-amplitude of the primary star was found to be $K_W = 
93 \pm 10$ km\,s$^{-1}$ from the motion of the wings of the HeII$\:
\lambda4686$\AA$\:$ emission line. By detecting weak absorption features 
from the secondary star, we find its radial velocity semi-amplitude to be 
$K_R = 170 \pm 10$ km\,s$^{-1}$. From these parameters, we obtain a mass of 
$M_1 = 1.55 \pm 0.24M_\odot$ for the white dwarf primary star and a mass of 
$M_2 = 0.88 \pm 0.17M_\odot$ for the secondary star. The radius of the 
secondary is calculated to be $R_2 = 2.1\pm0.2R_\odot$, confirming that 
it is evolved. The inclination of the system is calculated to be 
$i = 82.7^\circ\pm2.9^\circ$, consistent with the deep eclipse seen in the 
lightcurves. The helium emission lines are double-peaked, with the 
blue-shifted regions of the disc being eclipsed prior to the red-shifted 
regions, clearly indicating the presence of an accretion disc. The high mass 
of the white dwarf is consistent with the thermonuclear runaway model of 
recurrent nova outbursts, and confirms that U~Sco is the best Type Ia 
supernova progenitor currently known. We predict that U~Sco is likely to 
explode within $\sim 700\,000$ years.

\end{abstract} 


\section{Introduction}
\label{sec:introduction}

U~Sco belongs to a small class of cataclysmic variables (CVs) known as 
recurrent novae (RNe), which show repeated nova outbursts on timescales 
of decades. Of these, U~Sco has the shortest recurrence period ($\sim 8\:$y), 
with recorded outbursts in 1863, 1906, 1936, 1979, 1987 and 1999. It is 
likely that others have been missed due to its proximity to the 
ecliptic ($4^\circ$).

The RN class is further divided into three groups, depending on the nature 
of the secondary star (red giant, sub-giant or dwarf), each of which have 
different eruption mechanisms (see \pcite{warner95a} \& \pcite{webbink87} for 
reviews). U~Sco is the only object of the three in the sub-giant class 
which eclipses (\pcite{schaefer90}), and also shows evidence of secondary 
absorption lines (\pcite{hanes85}), making it the best candidate for the 
determination of system parameters. 

The outburst mechanism for the U~Sco sub-class of RNe is believed to be a 
modified version of the thermonuclear runaway (TNR) model of classical 
novae outbursts (e.g. \pcite{starrfield85}). 
In this model, the primary star builds up a layer of material, accreted 
from the secondary star, on its surface. The temperature and density 
at the base of the layer become sufficiently high for nuclear reactions 
to begin. 
After ignition, the temperature rises rapidly and the reaction rates run 
away, until the radiation pressure becomes high enough to eject most of 
the accreted material. This process occurs when the accreted layer reaches 
a critical mass (\pcite{truran86}), the value of which is a strongly 
decreasing function of the white dwarf mass. Therefore, to reduce the time 
interval between eruptions to those observed in the RNe, either the mass 
accretion rate ($\dot M$) or the white dwarf mass must be increased. 
For RNe, the white dwarf mass is the more important constraint because there 
is an upper limit to $\dot M$ above which the degeneracy on the surface of the 
white dwarf weakens, and no powerful eruption can occur. For this model 
to account for the short times between outbursts as seen in RNe, the 
white dwarf must have a mass close to the Chandrasekhar limit. 
A tight constraint on the white dwarf mass in U~Sco is hence theoretically 
important, as it will provide a direct observational test of the 
TNR model for this group of RNe.

The mass of the white dwarf in U~Sco is also important in terms of binary 
evolution and the role of RNe as Type Ia supernova progenitors. 
\scite{livio94} suggested, on the basis of the abundance determinations of 
the ejecta of RNe, that the mass of the envelope ejected during the outburst 
might be smaller than the amount of accreted material. In the case of high 
mass white dwarf systems, the primary could be pushed over the Chandrasekhar 
limit to produce a supernova explosion. A tight constraint on the white 
dwarf mass and the mass accretion rate allows the time to supernova to be 
calculated, which according to \scite{starrfield88}, could be as little as 
50\,000 years.

There have been two previous attempts to measure the mass of the 
white dwarf in U~Sco. \scite{johnston92} attempted to use low 
resolution spectra to determine the radial velocity semi-amplitude of 
both components by looking at the emission and absorption line shifts. 
They calculated a white dwarf mass of 0.23--0.60$M_\odot$, a result disputed 
by \scite{Duerbeck93}, who obtained 10 spectra with poor phase coverage, but 
were able to conclude a high primary mass, although their errors were large 
($M_{1} = 1.16\pm0.69M_\odot$).
\scite{schaefer95} re-phased the radial velocity measurements of 
\scite{johnston92} and \scite{Duerbeck93} using a new orbital period. 
They concluded that due to inconsistent $\gamma$ 
velocities, a phase shift in the emission lines and a large scatter 
in the radial velocity curves, no reliable white dwarf mass could be 
derived from the data. 

In addition to this, \scite{hachisu00b} sucessfully modelled the quiescent 
light curve of U~Sco with a white dwarf mass of $M_{1} \sim1.37M_\odot$, a 
secondary mass of $M_{2} = 0.8$--$2.0M_\odot$ and an orbital inclination 
of $\sim80^\circ$. Indirect support that the TNR model is 
the outburst mechanism in U~Sco has also been found by the observation of 
luminous supersoft X-ray emission after the 1999 outburst 
(\pcite{kahabka99}). They also constrain the white dwarf to be massive 
from the temperature of the optically thick supersoft component. 

In this paper, we present new measurements of the radial velocity 
semi-amplitude of the primary and secondary stars in U~Sco and hence 
a new determination of the mass of the white dwarf in this system. 

\section{Observations}
\label{sec:observations}

On the nights of 1999 April 15--19 we obtained 51 spectra of
the recurrent nova U~Sco with the 3.9m Anglo-Australian Telescope (AAT) in 
Siding Springs, Australia.
We covered one complete orbit of U~Sco during this time, including an 
eclipse -- a full journal of observations is given in Table~\ref{tab:journal}.
The exposures were all 1500\,s with about 50\,s dead-time for archiving of 
data. The setup comprised of the RGO Spectrometer + 250mm camera, the TEK 
$1024\times1024\;$ CCD chip and the 1200V grating, which gave a 
wavelength coverage of approximately 4620\AA--5435\AA $\:$ at 1.6\AA 
$\:$ (95\,km\,s$^{-1}$) resolution. We also took spectra of 20 class IV and V 
spectral type templates ranging from F0 -- K2, and the flux standard 
LTT\,9239 (\pcite{hamuy92}). The 1.5 
arcsec slit was orientated to cover a comparison star $\sim 1$ arcmin west 
of U~Sco in order to correct for slit losses. Arc spectra were 
taken between every U~Sco exposure to calibrate instrumental flexure.
The nights were all photometric, and the seeing varied between 
approximately 1.0 arcsec and 1.8 arcsec throughout the five nights.

\begin{figure}
\psfig{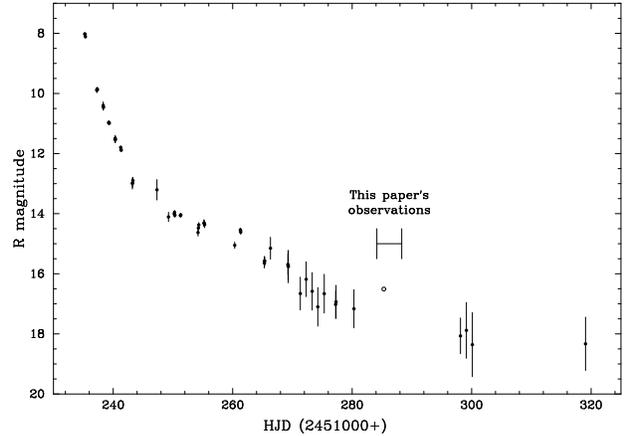}
\caption{R-band lightcurve of the 1999 outburst of U Sco. The data have been  
taken from \protect\scite{matsumoto01}. The open circle represents a 
measurement with an unrecorded error.}
\label{fig:outburst}
\end{figure}

\begin{table}
\caption{Journal of observations for U Sco. Orbital phase is calculated 
using the new ephemeris presented in this paper 
(equation 1).}
\label{tab:journal}
\begin{center}
\begin{tabular*}{84mm}{lr@{:}lr@{:}lcr@{.}lr@{.}l}
\hline
\vspace{0.5mm}\\
\multicolumn{1}{l}{UT Date} & \multicolumn{2}{c}{UT}
& \multicolumn{2}{c}{UT} & \multicolumn{1}{c}{No. of} 
& \multicolumn{2}{c}{Phase} & \multicolumn{2}{c}{Phase}\\
\multicolumn{1}{c}{} & \multicolumn{2}{c}{start}
& \multicolumn{2}{c}{end} & \multicolumn{1}{c}{spectra} 
& \multicolumn{2}{c}{start} & \multicolumn{2}{c}{end}
\vspace{4mm}\\
\hline
\vspace{0.5mm}\\
1999 April 15 & 14&57 & 19&09 & 8 & 0&31 & 0&44\\
1999 April 16 & 14&14 & 19&21 & 11 & 0&09 & 0&26\\
1999 April 17 & 14&21 & 19&11 & 10 & 0&92 & 1&07\\
1999 April 18 & 14&20 & 19&16 & 11 & 0&72 & 0&88\\
1999 April 19 & 14&23 & 19&30 & 11 & 0&54 & 0&70
\vspace{4mm}\\
\hline
\end{tabular*}
\end{center}
\end{table}

U~Sco underwent its sixth recorded outburst on 1999 February 25, 
reaching a maximum brightness of V = 7.6 (\pcite{schmeer99}). 
The observations recorded in this paper were taken approximately 49 days 
after outburst maximum, as can be seen from the lightcurve in 
Fig.~\ref{fig:outburst}, by which time the magnitude had decreased to 
approximately one magnitude above its quiescent level. Unfortunately, 
the single data point in Fig.~\ref{fig:outburst} corresponding to the dates 
of our observations has no recorded uncertainty. It is therefore 
impossible to conclude whether we actually observed U~Sco during an 
anomalously high state. What is certain, however, is that the additional shot 
noise in the continuum made it more difficult than we expected to detect 
absorption lines from the secondary star. The total flux in each spectrum is 
plotted in Fig.~\ref{fig:ephemeris} as a function of Heliocentric Julian Date 
(HJD) following equation~\ref{eqn:ephem} (see Section~\ref{sec:ephem}). 
The eclipse occurs on the third night of observation. 
There is also some evidence for a slight downward trend in brightness over 
the five nights.

\section{Data Reduction}
\label{sec:data reduction}

We debiased the data and corrected for the pixel-to-pixel variations with a 
tungsten lamp flat-field. The sky was subtracted by fitting second-order 
polynomials in the spatial direction to the sky regions 
on either side of the object spectra. The data were then optimally extracted 
(\pcite{horne86a}) to give raw spectra of U~Sco and the comparison star. 
Arc spectra were then extracted from the same locations on 
the detector as the targets. The wavelength scale for each spectrum 
was interpolated from the wavelength scales of two neighbouring arc 
spectra. The root-mean-square error in the fourth-order polynomial fits to 
the arc lines was $\sim0.025$\AA. \\

The next stage of the reduction process was to correct for the 
instrumental response and slit losses in order to obtain absolute 
fluxes. A third-order spline fit to the continuum of the 
spectrophotometric standard star was used to 
remove the large-scale variations of instrumental response with wavelength.
The slit-loss correction was performed by dividing the U~Sco spectra by 
spline fits to the comparison star spectra, and then multipling the resulting 
spectra by a spline fit to a wide-slit comparison star spectrum.

\section{Results}
\subsection{Ephemeris}
\label{sec:ephem}

\begin{figure}
\psfig{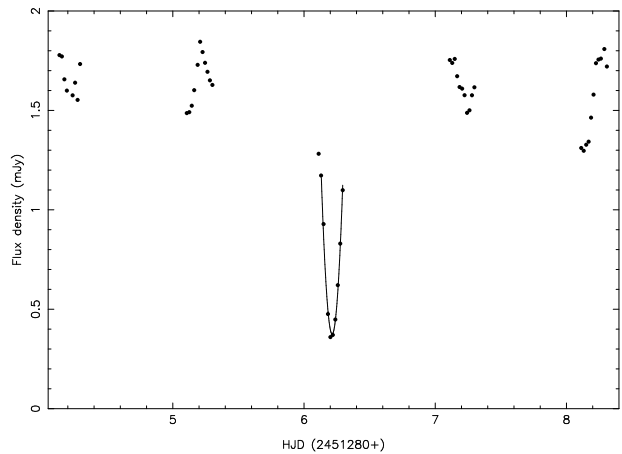}
\caption{Light curve of U~Sco during our observations. The solid curve is 
a parabolic fit to the eclipse minimum.}
\label{fig:ephemeris}
\end{figure}

The times of mid-eclipse were determined by fitting a parabola to 
the eclipse minimum (the solid curve in Fig.~\ref{fig:ephemeris}). A 
least-squares fit to the 9 eclipse timings found in \scite{schaefer95} 
and our eclipse yields the following ephemeris:

\begin{equation}
\label{eq:ephem}
\begin{array}{lrrl}
T_{\rm mid-eclipse} = & \!\!\!\!\! {\rm HJD}\,\,2\,447\,717.6145 
& \!\! + \,\, 1.230\,5522 & \!\!\!\!\! E \\
& \!\! \pm \,\, 0.0044 & \!\! \pm \,\, 0.000\,0024 & \\
\label{eqn:ephem}
\end{array}
\end{equation}

The differences between the observed and calculated times of mid-eclipse are 
given in Table~\ref{tab:o-cs}.  We find no evidence for a non-zero value 
for $\dot P$, in agreement with \scite{schaefer95}. 

\begin{table}
\caption{Times of mid-eclipse for U Sco according to Schaefer \& Ringwald 
(1995; SR95) and this paper.}
\label{tab:o-cs}
\begin{tabular*}{73mm}{@{}cr@{$\:\pm\:$}lr@{.}lc}
\hline
\vspace{0.5mm}\\
\multicolumn{1}{c}{Cycle} & \multicolumn{2}{c}{HJD at mid-eclipse}
& \multicolumn{2}{c}{O-C} & \multicolumn{1}{c}{Reference} \\
\multicolumn{1}{c}{(E)} & \multicolumn{2}{c}{(2,440,000+)}
& \multicolumn{2}{c}{(secs)} & \multicolumn{1}{c}{}
\vspace{4mm}\\
\hline
\vspace{0.5mm}\\
0 & 7717.5968&0.0139 & -1530&6 & SR95\\
1 & 7718.8412&0.0069 & -334&1 & SR95\\
4 & 7722.5318&0.0139 & -425&4 & SR95\\
5 & 7723.7783&0.0139 & 952&5 & SR95\\
8 & 7727.4699&0.0139 & 947&6 & SR95\\
1168 & 9154.9111&0.0104 & 1003&5 & SR95\\
1496 & 9558.5240&0.0069 & 293&1 & SR95\\
1497 & 9559.7497&0.0069 & -126&1 & SR95\\
1500 & 9563.4432&0.0139 & 33&2 & SR95\\
2900 & 11286.2143&0.0050 & -141&3 & This Paper
\vspace{4mm}\\
\hline
\end{tabular*}
\end{table}

\subsection{Average Spectrum}
\label{sec:average}

\begin{figure*}
\centerline{\psfig{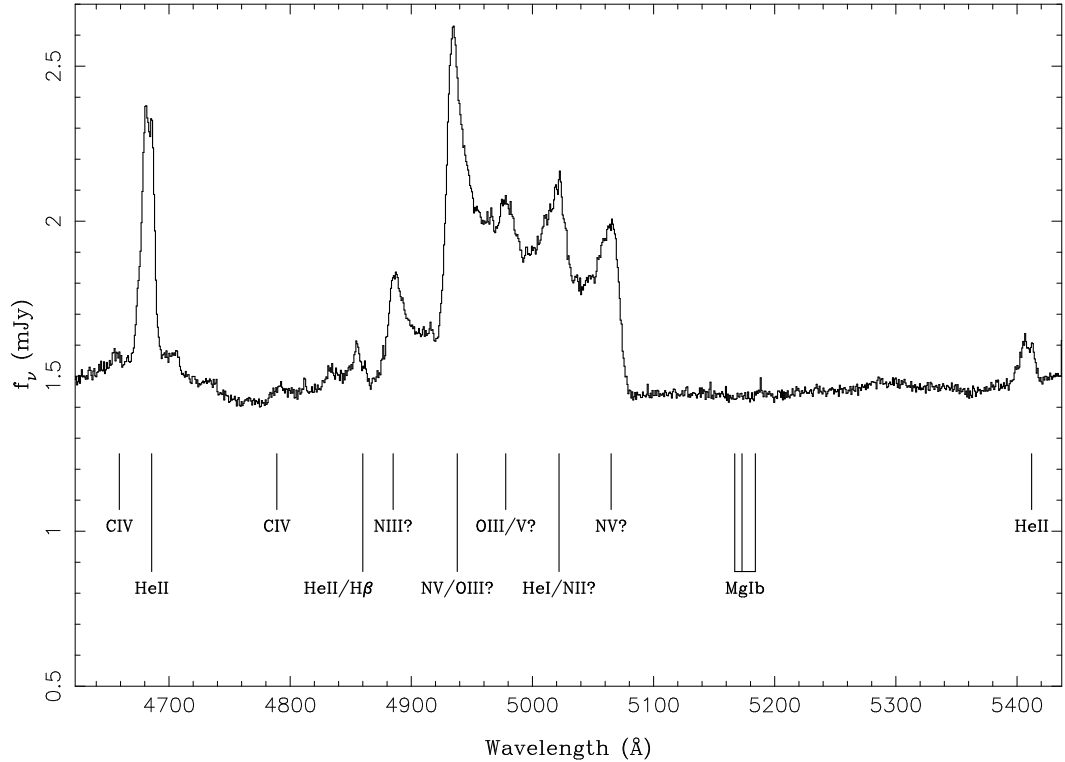}}
\caption{The average spectrum of U Sco, uncorrected for 
orbital motion.}
\label{fig:spectrum}
\end{figure*}

The average spectrum of U~Sco is displayed in Fig.~\ref{fig:spectrum}, and 
in Table~\ref{linewidths} we list fluxes, equivalent widths and velocity 
widths of the most prominent lines measured from the average spectrum.

The spectrum is dominated by single-peaked emission lines between 
$\lambda\lambda$4870--5080\AA$\:$ originating 
from the recent outburst. The identification of these nebular lines is 
complicated by their large widths, but the blend is likely to have 
contributions from NII, NIII, NV, OIII and OV, indicating a very high 
degree of excitation. 
Both \scite{sekiguchi88} and \scite{munari99} attribute the complex 
around $\lambda$5015\AA$\:$ to a blend of He I and NII in the 1987 and 
1999 outbursts, respectively. 
\scite{barlow81} suggest in the 1979 outburst the 
presence of H$\beta$ at $\lambda4861$\AA, and strong OIII emission at $\lambda
\lambda$4959/5007\AA. This is supported by comparison with nebulae around 
classical novae such as Nova Cygni 1992 (\pcite{moro01}). 
There is also evidence of CIV emission at $\lambda4658$\AA, in agreement with 
the 1987 outburst (\pcite{sekiguchi88}).
The other strong features in the blend are likely to be due to NIII and NV, 
in agreement with \scite{barlow81}, and optical spectra of Nova GQ Muscae 1983 
(\pcite{pequignot93}).
It is difficult to directly compare our spectra to the spectra of the 1979 
outburst (\pcite{barlow81}) and the 1987 outburst (\pcite{sekiguchi88}) 
because the latter two datasets were taken much closer to the eruption.

U~Sco also shows strong, broad, double-peaked He II lines at 
$\lambda$4686\AA$\;$ and $\lambda$5412\AA, typical of emission from an 
accretion disc in a high inclination system (e.g. \pcite{horne86}). 

The secondary star features cannot clearly be seen in 
Fig.~\ref{fig:spectrum}, as the spectra have been averaged without being 
corrected for orbital motion. Consequently, any faint absorption lines will 
have been smeared out. 

However, when the orbital motion of the secondary star is removed, the 
MgI{\em b} triplet ($\lambda\lambda$5167,5173,5184\AA) can clearly be 
identified (see Section~\ref{sec:secondary} and Figure~\ref{fig:templates}).


\begin{table*}
\caption[Fluxes and widths of prominent lines in U Sco]
{Fluxes and widths of prominent lines in U Sco, measured
from the average spectrum. The wavelength range of the blend is 
$\lambda\lambda$4870--5080\AA.}
\label{linewidths}
\begin{tabular*}{0.82\textwidth}{
lcr@{$\:\pm\:$}lcr@{$\:\pm\:$}lcr@{$\:\pm\:$}lcr@{$\:\pm\:$}l}
\hline
\vspace{0.5mm}\\
\multicolumn{1}{l}{Line} & \multicolumn{1}{c}{\hspace*{8mm}} & 
\multicolumn{2}{c}{Flux} & \multicolumn{1}{c}{\hspace*{8mm}} & 
\multicolumn{2}{c}{EW} & \multicolumn{1}{c}{\hspace*{8mm}} & 
\multicolumn{2}{c}{FWHM} & \multicolumn{1}{c}{\hspace*{8mm}} & 
\multicolumn{2}{c}{FWZI}\\
\multicolumn{1}{c}{} & & \multicolumn{2}{c}{$\times$ 10$^{-14}$}
& & \multicolumn{2}{c}{\AA} & & \multicolumn{2}{c}{km\,s$^{-1}$} 
& & \multicolumn{2}{c}{km\,s$^{-1}$}\\
\multicolumn{1}{c}{} & & \multicolumn{2}{c}{ergs\,cm$^{-2}$\,s$^{-1}$}
& & \multicolumn{2}{c}{} & & \multicolumn{2}{c}{} 
& &\multicolumn{2}{c}{}
\vspace{4mm}\\
\hline
\vspace{0.5mm}\\
HeII$\:\lambda$4686\AA & & 1.345&0.007 & & 6.98&0.04 & & 700&50 & & 
1600&100  \\

HeII$\:\lambda$5412\AA & & 0.177&0.005 & & 1.18&0.03 & & 700&100 & & 
1500&300  \\

Blend & & 11.95&0.02 & & 82.9&0.1 & & \multicolumn{2}{c}{} & & 
\multicolumn{2}{c}{} 
\vspace{4mm}\\
\hline
\end{tabular*}
\end{table*}

\subsection{Light Curves}
\label{sec:light}

The continuum light curve for U~Sco was computed by summing the flux in 
the wavelength range $\lambda\lambda$5080--5390\AA, which is devoid of 
emission lines. A polynomial fit to the continuum was then subtracted from the 
spectra and the emission line light curves were computed by summing the 
residual flux after continuum subtraction.

\begin{figure*}
\centerline{\psfig{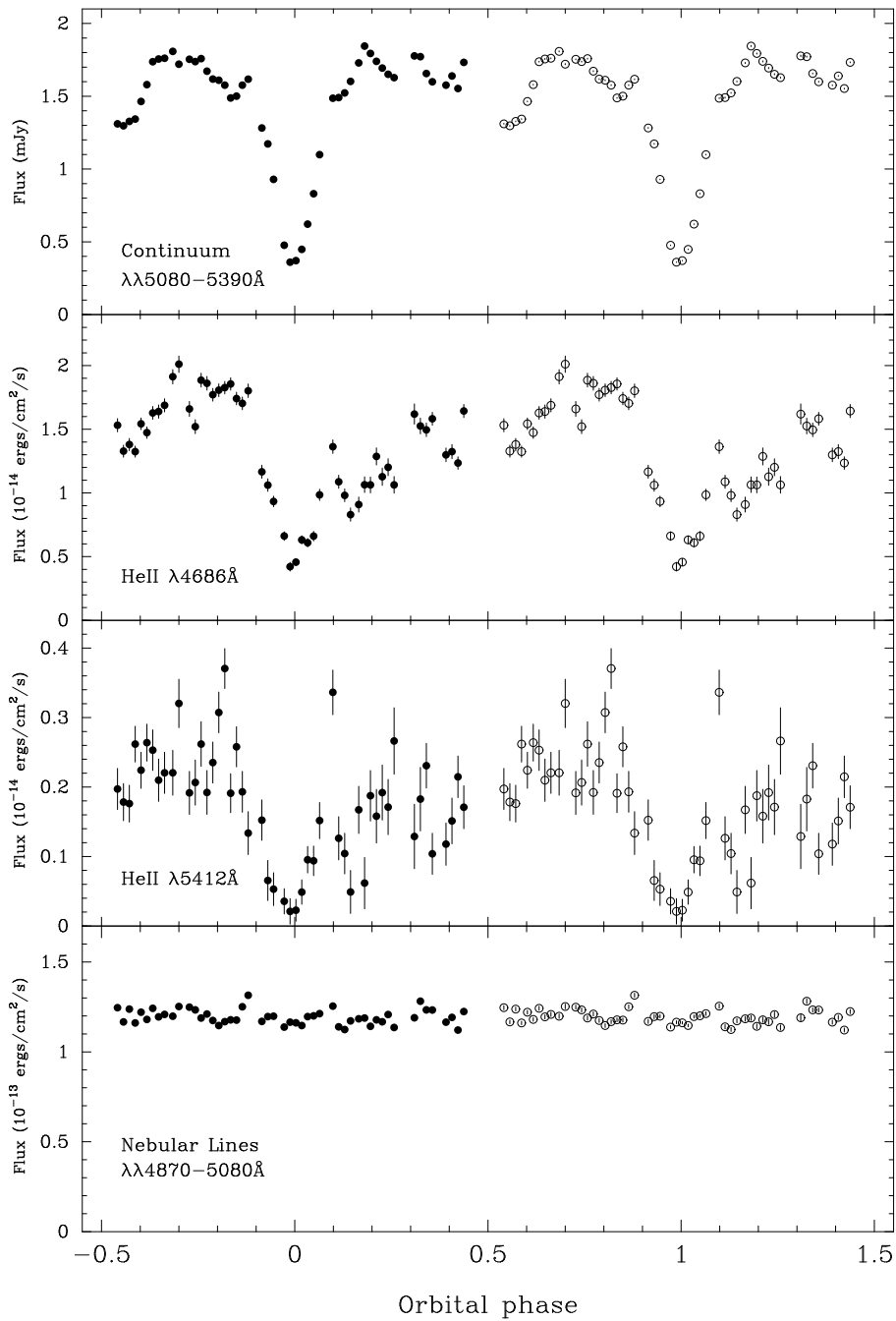}}
\caption{Continuum and emission line lightcurves of U Sco. The 
open circles represent points where the real data (the closed 
circles) have been folded over.}
\label{fig:lightcurve}
\end{figure*}

The resulting light curves are plotted in Fig.~\ref{fig:lightcurve} as 
a function of phase, following our new ephemeris. The continuum shows 
a symmetrical eclipse, with flickering throughout. There appears to 
be a dip in magnitude around phase 0.5, perhaps indicative of a secondary 
eclipse, but there are too few data points to be conclusive. 
In addition, as a consequence of U~Sco's long orbital period 
(1.23 days), the lightcurves represent data from five nights. Therefore, not 
all of the variability can be necessarily attributed to orbital effects.

The eclipses of the HeII lines have a different shape from the 
continuum. Both lightcurves show a long egress, and a pre-eclipse 
increase in brightness. There is no evidence for significant variability 
in the lightcurve of the nebular lines ($\lambda\lambda$4870--5080\AA), 
supporting the idea that the lines are produced in a region external to the 
orbits of the stellar components.

\subsection{Trailed spectrum}
\label{sec:trailed_spectrum}

\begin{figure*}
\centerline{\psfig{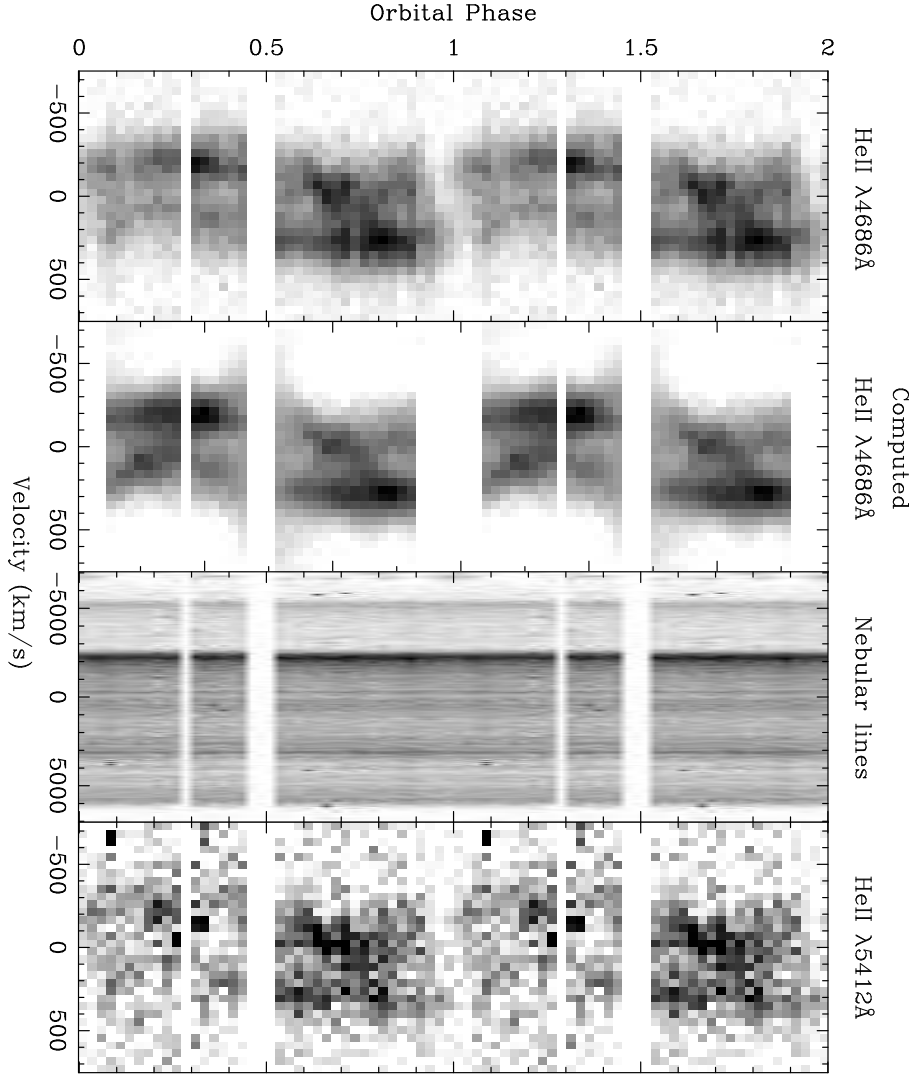}}
\caption{Trailed spectra of HeII $\lambda4686$\,\AA, the nebular 
lines and HeII $\lambda5412$\,\AA. 
The second panel from the left shows the HeII $\lambda4686$\,\AA$\:$ line 
computed from the Doppler map. 
The gaps around phases 0 and 1 in the computed data correspond to eclipse 
data, which were omitted from the fit; the 
gaps around phases 0.3, 0.5, 1.3 and 1.5 are due to 
missing data. 
The data only cover one cycle, but are folded over for clarity.}
\label{fig:ts}
\end{figure*}

We subtracted polynomial fits to the continua from the spectra and then 
rebinned the spectra onto a constant 
velocity-interval scale centred on the rest wavelength of the lines. 
The rest wavelength of the nebular lines was taken 
as $\lambda4975$\AA. The data were then phase binned into 40 bins, 36 of which 
were filled. Fig.~\ref{fig:ts} shows the trailed spectra of the 
HeII $\lambda4686$\AA, HeII $\lambda5412$\AA$\:$ and nebular lines in U~Sco.

The HeII lines show two peaks, which vary sinusoidally with phase. This, 
and the rotational disturbance (where the blue-shifted peak 
is eclipsed before the red-shifted peak) is evidence for origin in 
a high inclination accretion disc. There is also some evidence for an 
emission component moving from blue to red between phases 0.6 -- 0.9. 

\renewcommand{\textfraction}{0.3}

\subsection{Doppler Tomography}
\label{sec:doppler}

Doppler tomography is an indirect imaging technique which can be used 
to determine the velocity-space distribution of line emission in 
cataclysmic variables. For a detailed review of Doppler tomography, see 
\scite{marsh00}.

\begin{figure}
\psfig{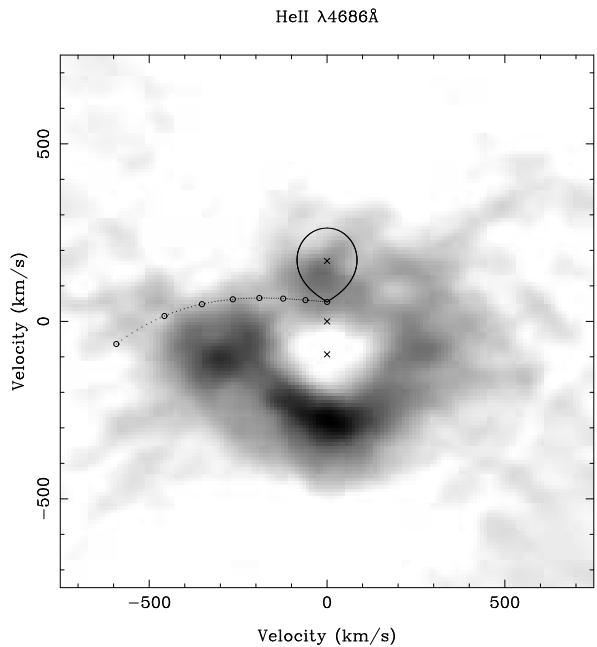}
\caption{Doppler map of HeII$\:\lambda4686$\AA$\:$ emission in U Sco.}
\label{fig:doppler}
\end{figure}

Fig.~\ref{fig:doppler} shows the Doppler map of the 
HeII$\;\lambda4686$\AA$\:$ line in U~Sco, computed from the trailed spectra of 
Fig.~\ref{fig:ts} but with the eclipse spectra removed. The data for 
the HeII$\:\lambda5412$\AA$\;$ line were too noisy to produce a Doppler map. 
The three crosses on the 
Doppler map represent the centre of mass of the secondary star (upper 
cross), the centre of mass of the system (middle cross) and the centre of 
mass of the white dwarf (lower cross). These crosses, the Roche lobe of the 
secondary star, and the predicted trajectory of the gas stream have been 
plotted using the radial velocities of the primary and secondary stars, 
$K_W = 93\pm10$ km\,s$^{-1}$ and $K_R = 170\pm10$ km\,s$^{-1}$, derived 
in Section~\ref{sec:params}. 
The series of circles along the gas stream mark the distance from the 
white dwarf at intervals of $0.1L_1$, ranging from $1.0L_1$ at the red star 
to $0.3L_1$.

A ring-like emission distribution, characteristic of a Keplerian accretion 
disc centred on the white dwarf, is clearly seen in Fig.~\ref{fig:doppler}.
There is some evidence for an increase in emission downstream from where 
the computed gas stream joins the accretion disc, implying the presence of 
a bright spot. This kind of behaviour has been seen in other CVs -- 
e.g. WZ Sge (\pcite{spruit98}).

\subsection{Radial velocity of the white dwarf}
\label{sec:whitedwarf}

\begin{figure}
\psfig{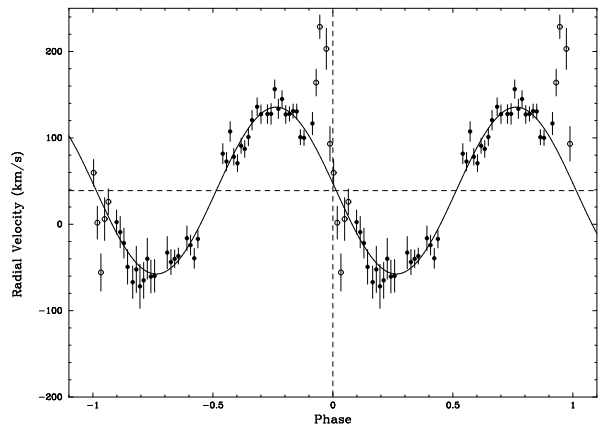}
\caption{Radial velocity curve of HeII $\lambda4686$\AA, measured 
using a double-Gaussian fit with a separation of 1000 km\,s$^{-1}$. 
Points marked by open circles were not included in the radial velocity 
fit, as they were affected by the rotational disturbance during primary 
eclipse. The horizontal dashed line represents the systemic velocity.}
\label{fig:radial}
\end{figure}

The continuum-subtracted data were binned onto a constant velocity 
interval scale about each of the two helium line rest wavelengths. 
In order to measure the velocities, we used the double-Gaussian method 
of \scite{schneider80}, since this technique is sensitive mainly to 
the motion of the line wings and should therefore reflect the motion 
of the white dwarf with the highest reliability. We varied the Gaussian 
widths between 150--300 km\,s$^{-1}$ at 50\,km\,s$^{-1}$ intervals, as well 
as varying their separation $a$ from 200--1500\,km\,s$^{-1}$. We then fitted
\begin{equation}
V=\gamma-K\sin[2\pi(\phi-\phi_0)]
\end{equation}
to each set of measurements, omitting the 9 points around primary eclipse 
which were affected by the rotational disturbance. 
An example of a radial velocity curve obtained for the 
HeII $\lambda4686$\AA$\;$ line for a Gaussian width of 300 km\,s$^{-1}$ and 
separation 1000 km\,s$^{-1}$ is shown in Fig.~\ref{fig:radial}. The data for 
the HeII $\lambda5412$\AA$\;$ line were too noisy to determine accurate 
radial velocities.

\begin{figure}
\psfig{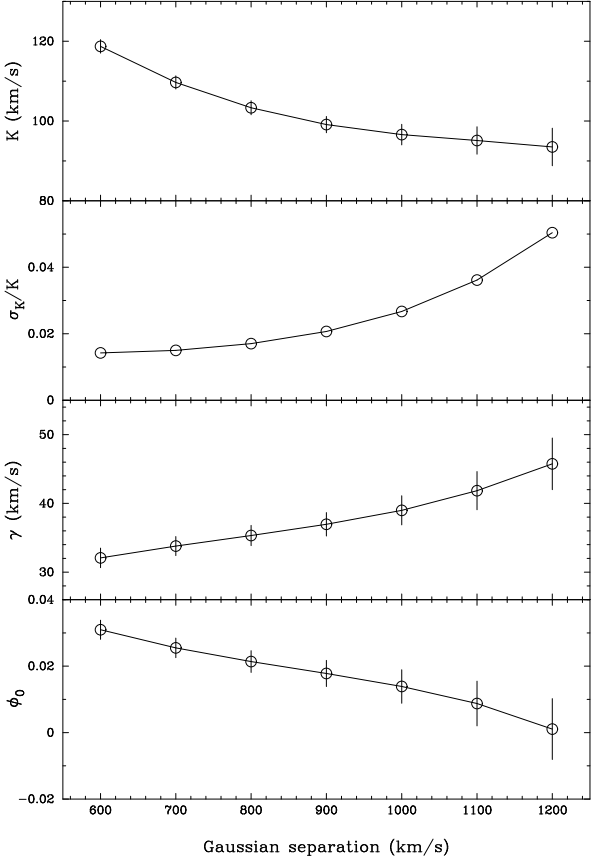}
\caption{The diagnostic diagram for U~Sco based on the double-Gaussian 
radial velocity fits to HeII $\lambda4686$\AA.}
\label{fig:diagnostic}
\end{figure}

The radial velocity curve has a negligible phase offset, an indication that 
the helium line is a reliable representation of the motion of the 
white dwarf. The results of the radial velocity analysis are displayed in 
the form of a diagnostic diagram in Fig.~\ref{fig:diagnostic}. 
By plotting $K$, its associated fractional error $\sigma_K/K$, $\gamma$ 
and $\phi_0$ as functions of the Gaussian separation, it is possible to 
select the value of $K$ that most closely represents the actual $K_W$ 
(\pcite{shafter86}). 
If the emission were disc dominated, one would expect the solution for $K$ to 
asymptotically reach $K_W$ when the Gaussian separation becomes sufficiently 
large, and furthermore, one would expect $\phi_0$ to fall to 0. This is 
seen to occur at a separation of $\sim$1200 km\,s$^{-1}$, corresponding to 
$K_W \sim\:$93 km\,s$^{-1}$. There is, however, no sudden increase in 
$\sigma_K/K$, prompting us to employ a different approach.

\begin{figure}
\psfig{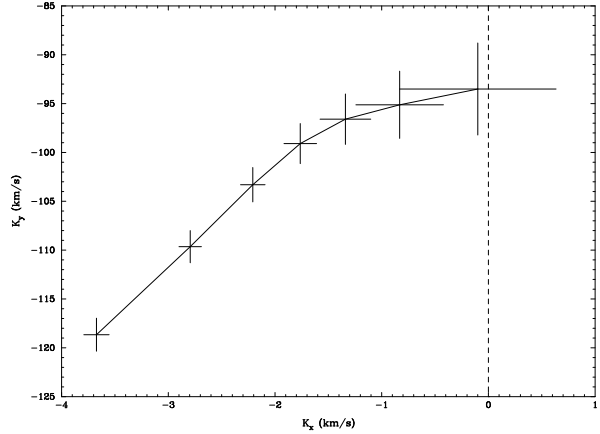}
\caption{Light centres diagram for HeII$\:\lambda4686$\AA$\:$ in U~Sco.}
\label{fig:centres}
\end{figure}

\scite{marsh88a} suggests that the use of a diagnostic diagram to evaluate 
$K_W$ does not account for systematic distortion of the radial velocity 
curve. We therefore attempted to make use of the light centres method, 
as described by \scite{marsh88a}. 
In the co-rotating co-ordinate system, 
the white dwarf has velocity ($0, -K_W$), and symmetric emission, say 
from a disc, would be centred at that point. By plotting 
$K_x = -K\sin\phi_0$ versus $K_y = -K\cos\phi_0$ for the different 
radial velocity fits (Fig.~\ref{fig:centres}), one finds that the points move 
closer to the $K_y$ axis with increasing Gaussian separation. A simple 
distortion which only affects low velocities, such as a bright spot, would 
result in this pattern, equivalent to a decrease in distortion as one measures 
emission closer to the velocity of the primary star. 
By extrapolating the last point on the light centre diagram to the $K_y$ 
axis, we measure the radial velocity semi-amplitude of the white dwarf 
$K_W = 93 \pm 10$ km\,s$^{-1}$. The error on $K_W$ was estimated from the 
uncertainty in crossing the $K_y$ axis, given the uncertainties associated 
with the points on the light centres diagram. Note the small scale on the 
$x$--axis of the light centres diagram.

\renewcommand{\textfraction}{0.2}

\subsection{Radial velocity of the secondary star}
\label{sec:secondary}

\begin{figure}
\psfig{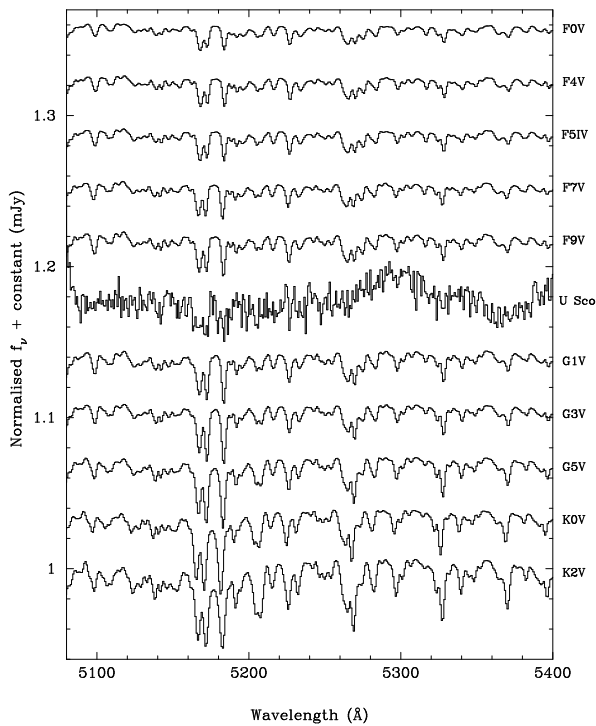}
\caption{Normalized spectra of ten template stars. The offset between each 
is 0.035. The U~Sco spectrum is a normalized average of 43 of the 51 spectra 
observed; 8 spectra were omitted due to cosmic rays interfering with 
secondary features. The template spectra have been broadened by 
16 km\,s$^{-1}$ to account for orbital smearing of the U~Sco spectra during 
exposure and by 88 km\,s$^{-1}$ to account for the rotational broadening 
of the lines in U~Sco.}
\label{fig:templates}
\end{figure}

The secondary star in U~Sco is observable through weak absorption lines, 
although many features have been drowned out by the nebular lines of the 
outburst. We compared regions of the spectra rich in absorption lines 
with several template dwarfs of spectral types F0--K2, the spectra 
of which are plotted in Fig.~\ref{fig:templates}. The absorption features are 
too weak for the normal technique of cross-correlation to be successful 
in finding the value of $K_R$, but it is possible to exploit these features 
to obtain an estimate of $K_R$ using the technique of skew mapping. 
This technique is descibed by \scite{smith93b}.

The first step was to shift the spectral type template stars 
to correct for their radial velocities. We then 
normalized each spectrum by dividing by a spline fit to the continuum and 
then subtracting 1 to set the continuum to zero. This ensures that line 
strength is preserved along the spectrum. The U~Sco spectra were also 
normalized in the same way. The template spectra were artificially 
broadened by 16 km\,s$^{-1}$ to account for orbital smearing of the U~Sco 
spectra through the 1500\,s exposures. The rotational velocity of the 
secondary star was found by the equation
\begin{equation}
q(1+q)^2 = 9.6 \Biggl({{v \sin i} \over {K_R}}\Biggr)^3
\end{equation}
(\pcite{smith98}) using estimated values of $q$ and $K_R$ in the first 
instance, then iterating to find the best fit values given in 
Section~\ref{sec:params}. The 
templates were then broadened by this value of $v\sin i$ = 88 km\,s$^{-1}$. 
Regions of the 
spectrum devoid of emission lines ($\lambda\lambda5085-5395$\AA) were then 
cross-correlated with each of the templates yielding a time series of 
cross-correlation functions (CCFs) for each template star.

\begin{figure}
\psfig{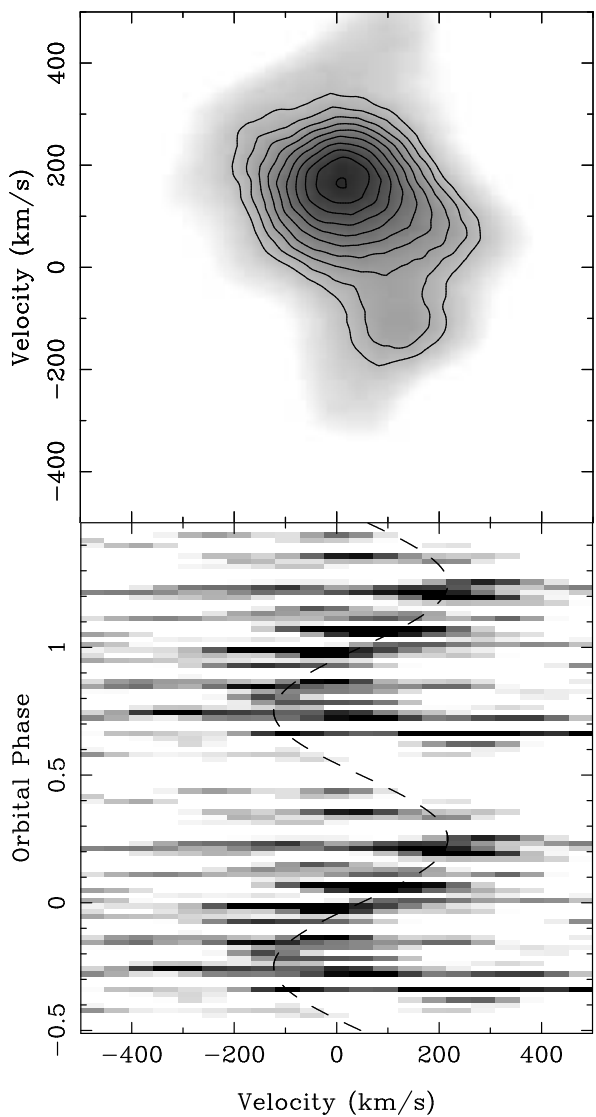}
\caption{The top panel shows the skew map of U~Sco for the K2V spectral type 
template. The bottom panel shows the trailed CCFs for the same 
template, clearly demonstrating that the peak in the skew map represents 
$K_R$ and is not an artefact of noise in the CCFs. The dashed line shows our 
adopted parameters for the secondary star orbit: $K_R = 170$ km\,s$^{-1}$, 
$\gamma = 47$ km\,s$^{-1}$ and zero phase shift. The data have been folded 
over for clarity.}
\label{fig:skewmap}
\end{figure}

To produce the skew map (Fig.~\ref{fig:skewmap}), these CCFs were 
back-projected in the same way as time-resolved spectra in standard 
Doppler tomography (\pcite{marsh88b}). If there is a detectable secondary 
star, we would expect a peak at (0,$K_R$) in the skew map. This can be 
repeated for each of the templates.

When we first back-projected the CCFs, the peak in each skew map was 
seen to be displaced to the blue of the $K_x = 0$ axis by around 
30 km\,s$^{-1}$. The reason was that we assumed 
that the centre of mass of the system was at rest. The systemic velocity of 
U~Sco derived from the radial velocity curve fits in 
Section~\ref{sec:whitedwarf} is $\gamma=47\pm10\:$km\,s$^{-1}$. 
Applying this $\gamma$ velocity shifts the peak to the red, towards the 
$K_x=0$ axis. As $K_x^2 \ll K_y^2$, then the value of $K_R$ 
is given by the equation
\begin{eqnarray}
 \nonumber
K_R = (K_x^2 + K_y^2)^{1/2} \approx K_y,
\end{eqnarray}
so $K_R$ changes little whilst varying $\gamma$. 

The skew maps produced using each of the template stars with 
$\gamma=0\:$km\,s$^{-1}$ show well-defined peaks at $(K_x,K_y) \sim 
(-30,160)\:$km\,s$^{-1}$. However, applying $\gamma=47\:$km\,s$^{-1}$, the 
peaks in the skew maps shift to $(K_x,K_y) \sim (15,180)\:$km\,s$^{-1}$. 
We suspect that the true $\gamma$ value of U~Sco falls between these two 
limits.
The skew map shown in Fig.~\ref{fig:skewmap} is for the K2V spectral type 
template with $\gamma$ = 47 km\,s$^{-1}$, and is arguably the best fitting 
template for the secondary star. The skew map shows a clear peak 
around $(K_x,K_y) \sim (10,165)\:$km\,s$^{-1}$. 
To bring the skew map peak to lie on $K_x = 0$, the $\gamma$ value must be 
modified to $\sim$ 30 km\,s$^{-1}$, which still falls within $2\sigma$ of 
the $\gamma$ derived from the emission lines. 
The resulting peak falls on $K_y = 170$ km\,s$^{-1}$, which we adopt as the 
radial velocity semi-amplitude for the secondary star: 
$K_R = 170 \pm10$ km\,s$^{-1}$. The uncertainty in $K_R$ has been 
determined from the scatter in $K_R$ due to four sources of error: 
varying the $\gamma$ velocity; using different spectral type 
templates; producing skew maps from random subsets of the data; the 
accuracy in measuring the peak in the skew map. 

The bottom panel of 
Fig.~\ref{fig:skewmap} shows a sine wave in the trailed CCFs, demonstrating 
that the peak in the skew map is not due to noise in the CCFs.
It can be seen that the peaks in the CCFs are most prominent around
primary eclipse ($\phi\sim$ 0.7--1.2). There are two possible
explanations for this. First, by considering the geometry of the secondary 
eclipse in U~Sco, we estimate that a maximum of 30 per cent of the visible 
surface of U~Sco will be obscured by the large accretion disc (see 
Section~\ref{sec:params}) between phases 0.35--0.65.  
We would therefore expect the peaks in the CCFs to weaken
considerably at these phases, which is exactly as observed in
Figure~\ref{fig:skewmap}. This explanation is further supported by
the tentative evidence for a secondary eclipse presented in the continuum 
light curve (Figure~\ref{fig:lightcurve}) and it does not affect the values 
for the masses we have derived in Section~\ref{sec:params}. Second, it is 
possible that the absorption features on the inner hemisphere of the secondary
star are weakened by  irradiation due to the recent outburst, which
would again explain the loss of the peaks in the CCFs around phase
0.5. If the latter explanation  were true, it is possible that we have
overestimated the value of $K_R$, which would in turn lead us to
overestimate $M_1$ (see Figure~\ref{fig:masses}). The only way we can
reliably account for any systematic effects due to irradiation would be
to obtain higher signal-to-noise observations during quiescence and so
measure $v\sin i$ during primary eclipse.

\subsection{Rotational velocity and spectral type of the secondary star}
\label{sec:rotational}

Mass estimates of CVs using emission line studies suffer from systematic 
errors, and have been thoroughly discussed by a number of authors 
(e.g. \pcite{marsh88a}). More reliable mass estimates can be obtained by 
using $K_R$ in conjunction with the rotational velocity of the secondary 
star, $v \sin i$. We attempted to determine $v \sin i$ using the 
procedure outlined in \scite{sad98b}, but the data were too noisy.

Previous estimates of the secondary 
spectral type have ranged from F5V to K2III. \scite{hanes85} suggests 
G0$\pm$5III-V, corroborated by \scite{webbink87} with a GIII subgiant. Recent 
estimates have agreed on the subgiant nature of the secondary. 
\scite{schaefer90} estimated the spectral type as G3-6 from the colours at 
minimum light, whereas \scite{johnston92} measured the observed ratio of 
calcium lines to H$\delta$ and found F8$\pm$2. Based on the detection of the 
MgI$b$ absorption in the 1979 outburst spectrum, \scite{kahabka99} find the 
secondary to be consistent with a low-mass subgiant secondary of spectral 
type K2. This is supported by \scite{anupama00}, who estimate the secondary 
to be a K2 subgiant based on the indices of the MgI$b$ absorption band and 
the FeI + CaI absorption feature in the late-decline spectrum of the 1999 
outburst. Unfortunately, as can be seen in Fig.~\ref{fig:templates}, it is 
impossible to confirm a secondary spectral type using our data, although the 
MgI$b$ complex is clearly present.

\subsection{System Parameters}
\label{sec:params}

The radial velocity curves show little or no phase shift, giving us 
confidence that our measurement of $K_W$ reflects the motion of the white 
dwarf. Hence, together with $K_R$, our newly 
derived period and a measurement of the eclipse full width at half-depth 
($\Delta\phi_{1/2}$), we can proceed to calculate accurate masses free of 
many of the assumptions which commonly plague CV mass determinations. Our 
white dwarf radial velocity semi-amplitude, $K_W = 93\pm10$ km\,s$^{-1}$ 
compares favourably with the previous measurement of 
$K_W = 95 \pm 50$ km\,s$^{-1}$ by \scite{barlow81} after the 1979 outburst. 
\scite{schaefer95} derived the values $K_W = 87 \pm 32$ km\,s$^{-1}$ and 
$K_R = 56 \pm 27$ km\,s$^{-1}$, but concluded that these results were 
unreliable due to inconsistent $\gamma$ velocities, a phase shift in the 
emission lines and a large scatter in the radial velocity curves.

In order to determine $\Delta\phi_{1/2}$, we estimated the levels of flux 
outside the eclipse (the principle source of the error) and at eclipse 
minimum, and then measured the full width of eclipse half way 
between these points. The eclipse full width at half-depth for U~Sco was 
measured to be $\Delta\phi_{1/2} = 0.104\pm0.01$, in good agreement with 
$\Delta\phi_{1/2} = 0.106\pm0.01$ cited by \scite{schaefer95}. 

We have opted for a Monte Carlo approach similar to that of \scite{horne93} 
to calculate the system parameters and their errors. For a given set of 
$K_W$, $K_R$, $\Delta\phi_{1/2}$ and $P$, other system parameters are 
calculated as follows.

\begin{figure*}
\psfig{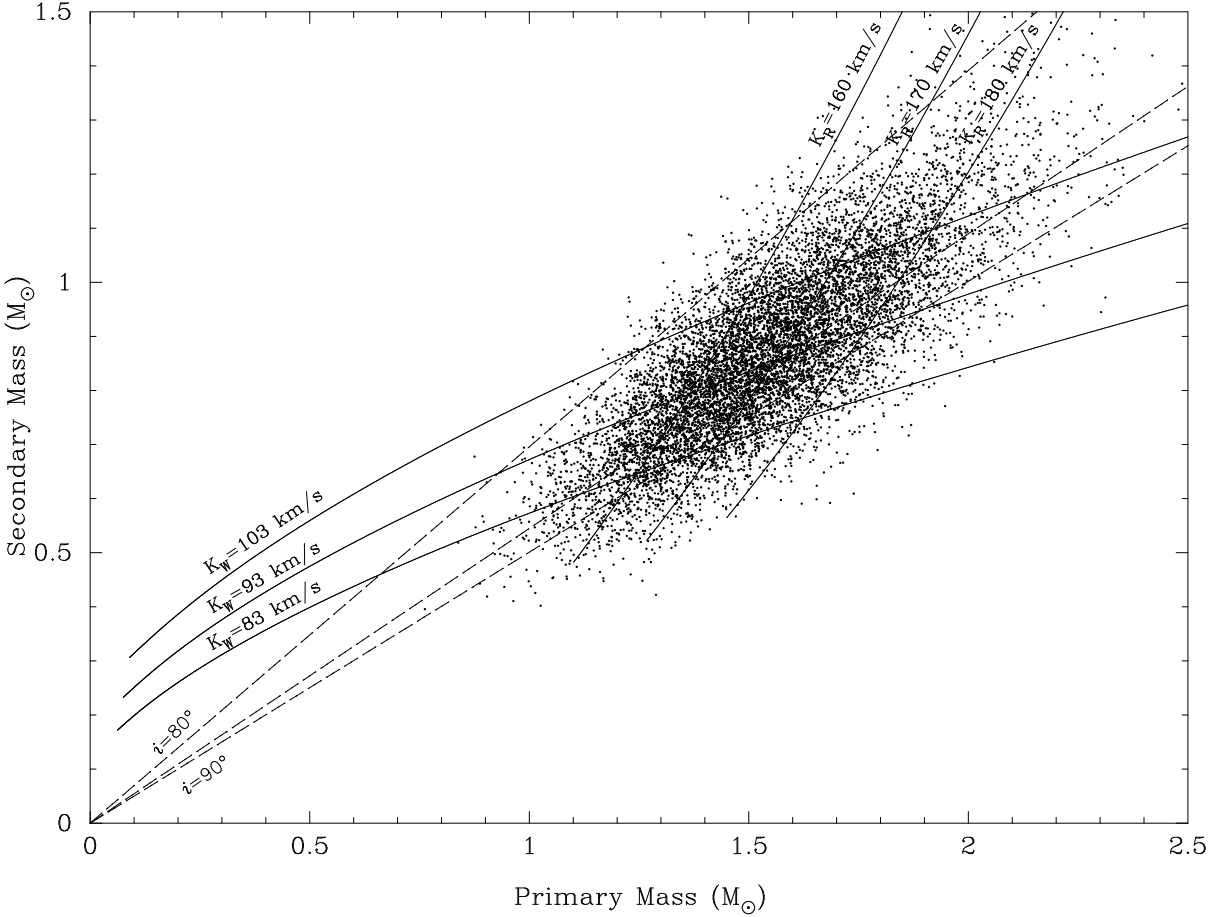}
\caption{Constraints on the masses of the stars in U~Sco. Each dot represents 
an ($M_1,M_2$) pair. The dashed lines are lines of constant inclination 
($i = 80^\circ$, $i = 85^\circ$, $i = 90^\circ$). The solid curves satisfy 
the constraints on $K_W = 93\pm10$\,km\,s$^{-1}$ and 
$K_R = 170\pm10$\,km\,s$^{-1}$.}
\label{fig:masses}
\end{figure*}

The mass ratio, $q$, can be determined from the ratio of 
the radial velocities
\begin{equation}
{q} = {M_1 \over M_2} = {K_W \over K_R}. 
\label{eqn:qratio}
\end{equation}
$R_2/a$ can be estimated because we know that the secondary star fills its 
Roche lobe (as there is an accretion disc present and hence mass transfer). 
$R_2$ is the equatorial radius of the secondary star and $a$ is the binary 
separation. We used Eggleton's formula (\pcite{eggleton83}) which gives the 
volume-equivalent radius of the Roche lobe to better than 1 per cent, which 
is close to the equatorial radius of the secondary star as seen during 
eclipse,
\begin{equation}
{{R_2} \over a} = {{0.49q^{2/3}} \over {{0.6q^{2/3} + \ln{(1+q^{1/3})}}}}. 
\label{eqn:eggleton}
\end{equation}
By considering the geometry of a point eclipse by a spherical body 
(e.g. \pcite{dhillon91}), the radius of the secondary can be shown to be
\begin{equation}
\biggl({{R_2} \over a}\biggr)^2 = \sin^2\pi\Delta\phi_{1/2}+
\cos^2\pi\Delta\phi_{1/2}\cos^2i.
\label{eqn:inclin}
\end{equation}
Using the value of $R_2/a$ obtained from equation~\ref{eqn:eggleton}, 
combined with equation~\ref{eqn:inclin}, the inclination of the system can 
be found. Kepler's Third Law gives
\begin{equation}
{{K_R^3P_{orb}}\over{2\pi G}}={{M_1\sin^3i}\over{(1+q)}^2}
\end{equation}
which, with the values of $q$ and $i$ calculated using 
equations~\ref{eqn:qratio}, \ref{eqn:eggleton}~\&~\ref{eqn:inclin}, gives 
the mass of the primary star. The mass of the secondary star can then be 
obtained using equation~\ref{eqn:qratio}.
The distance of each component to the centre of mass of the system 
($a_1$,$a_2$), is given by
\begin{equation}
{a_{(1,2)} = {{K_{(W,R)}P} \over {2\pi\sin{i}}}}
\label{eqn:asep}
\end{equation}
The separation of the two components, $a$, is $a_1 + a_2$, allowing the 
radius of the secondary star, $R_2$, to be calculated from 
equation~\ref{eqn:eggleton}.

The Monte Carlo simulation takes 10\,000 values of $K_W$, $K_R$, and 
$\Delta\phi_{1/2}$ (the error on the period is deemed to be negligible in 
comparison to the errors on $K_W$, $K_R$, and $\Delta\phi_{1/2}$), treating 
each as being normally distributed about their 
measured values with standard deviations equal to the errors on the 
measurements. We then calculate the masses of the components, the inclination 
of the system, the radius of the secondary star, and the separation of the 
components, as outlined above, omitting ($K_W$,$K_R$,$\Delta\phi_{1/2}$) 
triplets which are inconsistent with $\sin i \leq1$. Each accepted 
$M_1,M_2$ pair is then plotted as a point in Fig.~\ref{fig:masses}, and 
the masses and their errors are computed from the mean and standard deviation 
of the distribution of spots. The solid curves in Fig.~\ref{fig:masses} 
satisfy the white dwarf radial velocity constraint, 
$K_W = 93\pm10$\,km\,s$^{-1}$ and the secondary star radial velocity 
constraint, $K_R = 170\pm10$\,km\,s$^{-1}$. We find that 
$M_1 = 1.55 \pm 0.24M_\odot$ and $M_2 = 0.88 \pm 0.17M_\odot$. 
The inclination of the system is calculated to be 
$i = 82.7^\circ\pm2.9^\circ$, consistent with the nature of the deep eclipse 
seen in the lightcurves and the presence of double-peaked helium emission.

\begin{table}
\caption[System parameters for U~Sco]{System parameters for U~Sco}
\begin{tabular}{lr@{$\:\pm\:$}lr@{$\:\pm\:$}l}
\hline
\vspace{0.5mm}\\
\multicolumn{1}{l}{Parameter} &
\multicolumn{2}{c}{Measured Value} & 
\multicolumn{2}{c}{Monte Carlo Value}
\vspace{4mm}\\
\hline
\vspace{0.5mm}\\
$P_{orb}$ (d) & \multicolumn{2}{c}{1.2305522} & \multicolumn{2}{c}{} \\
$K_W$ (km\,s$^{-1}$) & 93&10 & 95&9 \\
$K_R$ (km\,s$^{-1}$) & 170&10 & 169&10 \\
$\Delta\phi_{1/2}$ & 0.104&0.010 & 0.098&0.007 \\
$q$ 		& \multicolumn{2}{c}{} & 0.55&0.07 \\
$i^\circ$ 	& \multicolumn{2}{c}{} & 82.7&2.9 \\
$M_1/M_\odot$	& \multicolumn{2}{c}{} & 1.55&0.24 \\
$M_2/M_\odot$ 	& \multicolumn{2}{c}{} & 0.88&0.17 \\
$R_2/R_\odot$ 	& \multicolumn{2}{c}{} & 2.1&0.2 \\
$a/R_\odot$ 	& \multicolumn{2}{c}{} & 6.5&0.4 \\
$\Delta\phi$	& 0.12&0.01 \\
$R_D/R_1$	& 0.87&0.11 
\vspace{4mm}\\
\hline
\end{tabular}
\label{tab:uscoparam}
\end{table}
We computed the radius of the accretion disc in U~Sco using the geometric 
method outlined in \scite{dhillon91}. The phase half-width of eclipse at 
maximum intensity was found to be $\Delta\phi = 0.12 \pm 0.01$ from 
Figure~\ref{fig:lightcurve}. Combining 
$\Delta\phi$ with $q$ and $i$ derived above produces an accretion disc 
radius of $R_D = 0.87 \pm 0.11 R_1$, where $R_1$ is the volume radius of the 
primary's Roche lobe. Our values for $\Delta\phi$ and $R_D$ are consistent 
with those found by \scite{harrop96}: $\Delta\phi = 0.129 \pm 0.008$ and 
$R_D \ge 0.92 R_1$.

The empirical relation obtained by \scite{smith98} between mass and radius 
for the secondary stars in CVs is given by,
\begin{equation}
\label{eq:smith}
{R \over R_\odot} = 
(0.93\pm0.09)\Biggl({M \over M_\odot}\Biggr)+(0.06\pm0.03).
\end{equation}
This predicts that if the secondary star is on the main-sequence, it should 
have a radius of 0.88$R_\odot$. The value we find is 
$2.1\pm0.2R_\odot$, clearly indicating that the white dwarf in U~Sco 
has an evolved companion. 

The values of all the system parameters of U Sco 
derived in this section are listed in Table~\ref{tab:uscoparam}.

\section{Discussion}
\label{sec:discussion}

\subsection{TNR Model}
\label{sec:tnr}

The outburst events in RNe could be the result of the TNR model, as seen in 
classical novae, or the disc instability model of dwarf novae. It is likely 
that given the heterogeneous nature of RNe, the mechanism depends on the 
individual system (\pcite{webbink90}). In the case of U~Sco, the TNR model was 
suggested in the light of severe problems with the disc instability model 
(\pcite{webbink87}). The motivation behind measuring the mass of the white 
dwarf in U~Sco was to give positive confirmation that outbursts are the 
result of a TNR, which predicts that the higher the mass of the white dwarf, 
the more frequent the outbursts. For eruptions to recur 
on the timescales seen in U~Sco ($\sim$ 8 years), the mass of the white 
dwarf must be very close to the Chandrasekhar mass of $1.378M_\odot$ 
(\pcite{starrfield85}; \pcite{starrfield88}; \pcite{nomoto84}).
Our observations allow us to calculate the mass of the white dwarf in 
U~Sco to be $M_1 = 1.55 \pm 0.24M_\odot$. Despite the large error on the 
value, we can conclude that U~Sco contains a high mass white dwarf, 
although additional work, such as an accurate $v \sin i$ determination during 
quiescence, must be done to further constrain this.

\subsection{U~Sco as a Type Ia SN Progenitor}
\label{sec:super}

Type Ia supernovae (SNe Ia) are widely believed to be thermonuclear 
explosions of mass-accreting white dwarfs (see \pcite{nomoto97} for a 
review). The favoured model for progenitor binary systems is the 
Chandrasekhar mass model, in which a mass-accreting carbon-oxygen white 
dwarf grows in mass up to the Chandrasekar limit and then explodes as a SN 
Ia. For the evolution of accreting WDs towards the Chandrasekhar mass, two 
scenarios have been proposed. The first is a double degenerate scenario, 
where two white dwarfs merge to cross the Chandrasekhar limit. 
A candidate for this has been identified in KPD 1930+2752 (\pcite{maxted01}). 
The second 
is a single degenerate scenario, where matter is accreted via mass transfer 
from a stellar companion. A new evolutionary path for single degenerate 
progenitor systems is discussed by \scite{hachisu99}, which descibes how 
the secondary (a slightly evolved main-sequence star) becomes helium rich. 
Their progenitor model predicts helium enriched matter accretion onto 
a white dwarf. Objects which display these characteristics in the form of 
strong HeII$\:\lambda4686$\AA$\:$ lines are luminous super-soft X-ray sources 
and certain recurrent novae, such as U~Sco and the others in the U~Sco 
subclass (V394 CrA and LMC-RN). The observations in this paper, in 
confirming the helium rich accreted matter, and constraining the white 
dwarf mass to be near the Chandrasekhar mass, make U~Sco the best candidate 
for this evolutionary path to a SN Ia.

The time to supernova can be estimated by dividing the amount of 
mass required to reach the Chandrasekhar mass by the average mass accretion 
rate ($\dot M_{av}$). Our results suggest a minimum white dwarf mass of 
$\sim1.31 M_\odot$, which requires it to accrete approximately 0.07$M_\odot$ 
to become a SN Ia. In their model, \scite{hachisu00a} find that the white 
dwarf in U~Sco grows in mass at an average rate of $\dot M_{av} 
\sim 1.0 \times 10^{-7} M_\odot$ y$^{-1}$. Using this value, we 
therefore predict U~Sco to become a supernova within $\sim700\,000$ years.

\section{Conclusions}
\label{sec:conclusions}

\begin{enumerate}

\item{We have shown that U~Sco contains a high mass white dwarf 
($M_1 = 1.55 \pm 0.24M_\odot$), confirming that the outburst mechanism is 
the TNR model.}

\item{The presence of an accretion disc has been confirmed for the first 
time by clear evidence of rotational disturbance in the HeII emission lines 
during eclipse, as well as their double-peaked nature.}

\item{The secondary star has a radius of $2.1\pm0.2R_\odot$, consistent 
with the idea that it is evolved.}

\item{The mass of the white dwarf in U~Sco is $M_1 = 1.55 \pm 0.24M_\odot$, 
implying that it is the best SN Ia candidate known and is expected to explode 
in $\sim$ 700\,000 years.}

\end{enumerate}

\section*{\sc Acknowledgements}

We would like to thank Katsura Matsumoto, Taichi Kato and Izumi Hachisu for 
their light curve of U~Sco during the 1999 outburst, and Chris Watson 
for useful comments.
TDT and SPL are supported by PPARC studentships. 
The authors acknowledge the data analysis facilities at Sheffield
provided by the Starlink Project which is run by CCLRC on behalf of PPARC. 
The Anglo-Australian Telescope is operated at Siding Springs by the AAO.

\bibliographystyle{mnras}
\bibliography{refs}

\begin{thebibliography}{{Moro-Mart\'{i}n, Garnavich \& Noriega-Crespo}{2001}}

\bibitem[\protect\citefmt{Anupama \& Dewangan}{2000}]{anupama00}
Anupama~G.~C., Dewangan~G.~C., 2000, AJ, 119, 1359

\bibitem[\protect\citefmt{Barlow {\rm et~al.}}{1981}]{barlow81}
Barlow~M.~J. {\rm et~al.}, 1981, MNRAS, 195, 61

\bibitem[\protect\citefmt{Dhillon, Marsh \& Jones}{1991}]{dhillon91}
Dhillon~V.~S., Marsh~T.~R., Jones~D. H.~P., 1991, MNRAS, 252, 342

\bibitem[\protect\citefmt{Duerbeck {\rm et~al.}}{1993}]{Duerbeck93}
Duerbeck {\rm et~al.}, 1993, ESO Messenger, 71, 19

\bibitem[\protect\citefmt{Eggleton}{1983}]{eggleton83}
Eggleton~P.~P., 1983, ApJ, 268, 368

\bibitem[\protect\citefmt{Hachisu {\rm et~al.}}{1999}]{hachisu99}
Hachisu~I., Kato~M., Nomoto~K., Umeda~H., 1999, ApJ, 519, 314

\bibitem[\protect\citefmt{Hachisu {\rm et~al.}}{2000a}]{hachisu00b}
Hachisu~I., Kato~M., Kato~T., Matsumoto~K., Nomoto~K., 2000a, ApJ, 534, L189

\bibitem[\protect\citefmt{Hachisu {\rm et~al.}}{2000b}]{hachisu00a}
Hachisu~I., Kato~M., Kato~T., Matsumoto~K., 2000b, ApJ, 528, L97

\bibitem[\protect\citefmt{Hamuy {\rm et~al.}}{1992}]{hamuy92}
Hamuy~M., Walker~A.~R., Suntzeff~N.~B., Gigoux~P., Heathcote~S.~R.,
  Phillips~M.~M., 1992, PASP, 104, 677

\bibitem[\protect\citefmt{Hanes}{1985}]{hanes85}
Hanes~D.~A., 1985, MNRAS, 213, 443

\bibitem[\protect\citefmt{Harrop-Allin \& Warner}{1996}]{harrop96}
Harrop-Allin~M.~K., Warner~B., 1996, MNRAS, 279, 219

\bibitem[\protect\citefmt{Horne \& Marsh}{1986}]{horne86}
Horne~K., Marsh~T.~R., 1986, MNRAS, 218, 761

\bibitem[\protect\citefmt{Horne, Welsh \& Wade}{1993}]{horne93}
Horne~K., Welsh~W.~F., Wade~R.~A., 1993, ApJ, 410, 357

\bibitem[\protect\citefmt{{Horne}}{1986}]{horne86a}
{Horne}~K., 1986, PASP, 98, 609

\bibitem[\protect\citefmt{Johnston \& Kulkarni}{1992}]{johnston92}
Johnston~H.~M., Kulkarni~S.~R., 1992, ApJ, 396, 267

\bibitem[\protect\citefmt{Kahabka {\rm et~al.}}{1999}]{kahabka99}
Kahabka~P., Hartmann~H.~W., Parmar~A.~N., Negueruela~I., 1999, AA, 347, L43

\bibitem[\protect\citefmt{Livio \& Truran}{1994}]{livio94}
Livio~M., Truran~J.~W., 1994, ApJ, 425, 797

\bibitem[\protect\citefmt{Marsh \& Horne}{1988}]{marsh88b}
Marsh~T.~R., Horne~K., 1988, MNRAS, 235, 269

\bibitem[\protect\citefmt{Marsh}{1988}]{marsh88a}
Marsh~T.~R., 1988, MNRAS, 231, 1117

\bibitem[\protect\citefmt{Marsh}{2000}]{marsh00}
Marsh~T.~R., 2000, in Boffin~H., Steeghs~D., eds, Proceedings of the
  International Workshop on Astro-tomography, Brussels, July 2000.
\newblock Springer-Verlag Lecture Notes in Physics, in press

\bibitem[\protect\citefmt{Matsumoto, Kato \& Hachisu}{2001}]{matsumoto01}
Matsumoto~K., Kato~K., Hachisu~I., 2001, PASP, submitted

\bibitem[\protect\citefmt{Maxted, Marsh \& North}{2000}]{maxted01}
Maxted~P. F.~L., Marsh~T.~R., North~R.~C., 2000, MNRAS, 317, L41

\bibitem[\protect\citefmt{Moro-Mart\'{i}n, Garnavich \&
  Noriega-Crespo}{2001}]{moro01}
Moro-Mart\'{i}n~A., Garnavich~P.~M., Noriega-Crespo~A., 2001, AJ, 121, 1636

\bibitem[\protect\citefmt{Munari {\rm et~al.}}{1999}]{munari99}
Munari~U. {\rm et~al.}, 1999, AA, 347, L39

\bibitem[\protect\citefmt{Nomoto, Iwamoto \& Kishimoto}{1997}]{nomoto97}
Nomoto~K., Iwamoto~K., Kishimoto~N., 1997, Sci, 276, 1378

\bibitem[\protect\citefmt{Nomoto, Thielemann \& Yokoi}{1984}]{nomoto84}
Nomoto~K., Thielemann~F., Yokoi~K., 1984, ApJ, 286, 644

\bibitem[\protect\citefmt{P\'{e}quignot {\rm et~al.}}{1993}]{pequignot93}
P\'{e}quignot~D., Petitjean~P., Boisson~C., Krautter~J., 1993, AA, 271, 219

\bibitem[\protect\citefmt{Schaefer \& Ringwald}{1995}]{schaefer95}
Schaefer~B.~E., Ringwald~F.~A., 1995, ApJ, 447, L45

\bibitem[\protect\citefmt{Schaefer}{1990}]{schaefer90}
Schaefer~B.~E., 1990, ApJ, 355, L39

\bibitem[\protect\citefmt{Schmeer {\rm et~al.}}{1999}]{schmeer99}
Schmeer~P., Waagen~E., Shaw~L., Mattiazzo~M., 1999, IAU Circ., 7113, 1

\bibitem[\protect\citefmt{Schneider \& Young}{1980}]{schneider80}
Schneider~D.~P., Young~P.~J., 1980, ApJ, 238, 946

\bibitem[\protect\citefmt{Sekiguchi {\rm et~al.}}{1988}]{sekiguchi88}
Sekiguchi~M.~W., Feast~M.~W., Whitelock~P.~A., Overbeek~M.~D., Wargau~W.,
  Spencer~Jones~J., 1988, MNRAS, 234, 281

\bibitem[\protect\citefmt{Shafter, Szkody \& Thorstensen}{1986}]{shafter86}
Shafter~A.~W., Szkody~P., Thorstensen~J.~R., 1986, ApJ, 308, 765

\bibitem[\protect\citefmt{Smith \& Dhillon}{1998}]{smith98}
Smith~D.~A., Dhillon~V.~S., 1998, MNRAS, 301, 767

\bibitem[\protect\citefmt{Smith, Cameron \& Tucknott}{1993}]{smith93b}
Smith~R.~C., Cameron~A., Tucknott~D.~S., 1993, in Regev~O., Shaviv~G., eds,
  Cataclysmic Variables and Related Physics.
\newblock Inst. Phys. Publ., Bristol, p.~70

\bibitem[\protect\citefmt{{Smith}, {Dhillon} \& {Marsh}}{1998}]{sad98b}
{Smith}~D.~A., {Dhillon}~V.~S., {Marsh}~T.~R., 1998, MNRAS, 296, 465

\bibitem[\protect\citefmt{Spruit \& Rutten}{1998}]{spruit98}
Spruit~H.~C., Rutten~R. G.~M., 1998, MNRAS, 299, 768

\bibitem[\protect\citefmt{Starrfield, Sparks \& Shaviv}{1988}]{starrfield88}
Starrfield~S., Sparks~W.~M., Shaviv~G., 1988, ApJ, 325, L35

\bibitem[\protect\citefmt{Starrfield, Sparks \& Truran}{1985}]{starrfield85}
Starrfield~S., Sparks~W.~M., Truran~J.~W., 1985, ApJ, 291, 136

\bibitem[\protect\citefmt{Truran \& Livio}{1986}]{truran86}
Truran~J.~W., Livio~M., 1986, ApJ, 308, 721

\bibitem[\protect\citefmt{Warner}{1995}]{warner95a}
Warner~B., 1995, Cataclysmic Variable Stars.
\newblock Cambridge University Press, Cambridge

\bibitem[\protect\citefmt{Webbink {\rm et~al.}}{1987}]{webbink87}
Webbink~R.~F., Livio~M., Truran~J.~W., Orio~M., 1987, ApJ, 314, 653

\bibitem[\protect\citefmt{Webbink}{1990}]{webbink90}
Webbink~R.~F., 1990, in Cassatella~A., Viotti~R., eds, Proceedings of
  Colloquium No. 122 of the International Astronomical Union, Madrid, June
  1989.
\newblock Springer-Verlag, Berlin, p.~405

\end{thebibliography}

\end{document}